\documentclass[11pt, a4paper,onecolumn]{belarticle4}
\pdfoutput=1
\usepackage{revsymb, amsmath, amsfonts, amssymb, enumerate, fullpage, amsthm, graphicx, braket, relsize, bbm, mathrsfs, mathtools}
\usepackage[normalem]{ulem}
\usepackage[T1]{fontenc}

\usepackage{graphicx,color}
\usepackage{paralist}
\usepackage{support-caption}

\usepackage[numbers,sort&compress]{natbib}
\usepackage{xfrac}

\usepackage[linktocpage=true, colorlinks=true, linkcolor=red, urlcolor=blue, citecolor=blue]{hyperref}

\newcommand{\cH}{\mathcal{H}}

\newcommand{\id}{\mathbb{I}}

\newcommand{\tr}[2]{\mathrm{tr}_{#2} \left\{ #1 \right\}}
\newcommand{\Tr}[1]{\mathrm{tr}\left\{#1 \right\}}

\newcommand{\av}[1]{\langle #1 \rangle}

\newcommand{\X}{\mathbb{X}}

\newcommand{\A}{\mathbb{A}}
\newcommand{\As}{\boldsymbol{\Sigma}}

\begin{document}

\title{Robust self-testing of quantum steering assemblages via operator inequalities}

\author{Beata Zjawin}
\affiliation{International Centre for Theory of Quantum Technologies, University of  Gda{\'n}sk, 80-308 Gda{\'n}sk, Poland}
\email{beata.zjawin@ug.edu.pl}


\begin{abstract}
Robust self-testing provides a framework for certifying quantum resources under experimental imperfections. Improving robustness bounds for quantum resources such as quantum states, steering assemblages, and measurements is a constant effort that ensures relevance in the experimental realm. Despite progress in analytic self-testing methods for quantum states and measurements, extending these techniques to device-independent certification of steering assemblages has remained an open challenge, with previous work relying primarily on numerical approaches. We address this gap by developing operator inequalities for robust self-testing of quantum steering assemblages. Specifically, we consider the assemblage that achieves maximal violation of the Clauser-Horne-Shimony-Holt (CHSH) inequality and obtain explicit lower bounds on its certification fidelity. Our analytic approach yields results that significantly improve upon previous numerical bounds, representing the first analytic treatment of device-independent assemblage self-testing. This work demonstrates a new application of operator inequalities beyond quantum state certification and contributes to the foundational understanding of device-independent certification in steering scenarios, potentially guiding future theoretical and experimental developments.
\end{abstract}

\maketitle
%

%
%
\newpage

\section{Introduction}

Correlations that arise from incompatible measurements on systems prepared in entangled states cannot always be explained by classical models. Bell's theorem~\cite{bell1964einstein,bell1966problem} formalizes this observation and provides methods to certify it through violations of Bell inequalities. Remarkably, certain quantum correlations can arise only from specific physical realizations. This connection between quantum resource structure and experimental statistics has motivated the development of self-testing~\cite{MY98,MY04,Bell-Nonlocality,vsupic2020self} -- a framework for identifying which quantum correlations uniquely determine the underlying physical setup. To accommodate experimental imperfections and noise, this framework has been generalized to robust self-testing. The goal in this setting is to show that if the observed correlations are close to the ideal ones, then the underlying quantum resource must be close to the reference resource being certified.

Although self-testing predominantly focuses on certifying quantum states and measurements, the framework naturally extends to other quantum resources. In this work, we focus on the steering scenario~\cite{einstein1935can,schrodinger1935discussion,cavalcanti2009experimental}, in which Alice and Bob share a bipartite quantum system and Alice performs local measurements on her subsystem. The key object of study is the \textit{assemblage}~\cite{pusey2013negativity}, a set of subnormalized states describing Bob’s system conditioned on Alice’s measurement choices and outcomes. Steering assemblages have many applications in quantum cryptography, such as one-sided device-independent quantum key distribution~\cite{branciard2012one}, randomness certification~\cite{passaro2015optimal} and randomness expansion~\cite{skrzypczyk2018maximal}. From the foundational perspective, steering is closely connected to measurement incompatibility~\cite{quintino2014joint,uola2015one} and contextuality~\cite{tavakoli2020measurement,plavala2022incompatibility}, further motivating its study as a resource of non-classicality~\cite{gallego2015resource,EPRLOSR}. 

The important role of steering in quantum information processing protocols motivates the development of robust self-testing methods for steering assemblages. While steering scenarios have long been known to be useful for self-testing quantum states~\cite{supic2016,gheorghiu2017rigidity}, robust self-testing focused on steering assemblages has only recently been explored~\cite{Chen2021robustselftestingof}. A key reason to focus on assemblages, rather than the underlying states and measurements, is that different quantum states and measurement sets -- even those not related by unitary transformations -- can produce the same assemblage. Moreover, Ref.~\cite{Chen2021robustselftestingof} shows that self-testing of assemblages provides an alternative approach for the device-independent certification of all entangled two-qubit states, as well as for the verification of all non-entanglement-breaking qubit channels. This approach is also central to recent protocols aimed at activating post-quantum steering~\cite{sainz2024activation,zjawin2024activation}. In Ref.~\cite{Chen2021robustselftestingof}, robust self-testing of steering assemblages is formally defined, and the authors present a semidefinite program to compute a lower bound on the fidelity between the underlying assemblage and the reference assemblage. However, obtaining analytic robust self-testing statements is left as an open problem.

In this work, we focus on the assemblage that maximally violates the Clauser-Horne-Shimony-Holt (CHSH) inequality \cite{clauser1969proposed}, which we refer to as the \textit{CHSH-type assemblage}. It is a crucial resource for many cryptographic tasks~\cite{branciard2012one,passaro2015optimal}. We prove the first analytic robust device-independent self-testing result for this assemblage. Our approach is based on the method of operator inequalities, originally developed for robust self-testing of quantum states~\cite{kaniewski2016analytic}. Central to this framework is the notion of extractability~\cite{bardyn2009device,kaniewski2016analytic}, which quantifies how well a reference quantum resource can be extracted from an unknown one. This method has provided the tightest known analytic bounds for the self-testing of the maximally entangled two-qubit state using the CHSH inequality, and the tripartite GHZ state using the Mermin inequality~\cite{kaniewski2016analytic}. It has also been extended to general two-qubit states~\cite{coopmans2019robust}, and adapted to the prepare-and-measure scenario~\cite{tavakoli2018self}. Here, we extend the operator inequality framework to steering scenarios and derive an analytic lower bound on the fidelity between the underlying assemblage and the CHSH-type assemblage, as a function of the observed CHSH violation. Our result significantly improves upon the numerical bound established in Ref.~\cite{Chen2021robustselftestingof}.

\section{Preliminaries} \label{sec:preliminaries}

\subsection{Quantum steering}

\begin{figure}[htbp]
  \centering
  \includegraphics[width=\linewidth]{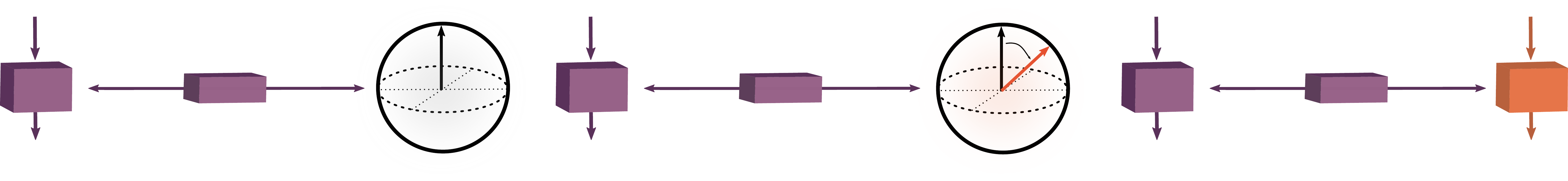}
\put(-395,25){$\rho_{AB}$}
\put(-397,-17){$(a)$}
\put(-454,8){$a$}
\put(-454,40){$x$}
\put(-335,54){$\rho_{a|x}$}
\put(-235,25){$\rho_{AB}$}
\put(-233,-17){$(b)$}
\put(-292,8){$a$}
\put(-292,40){$x$}
\put(-180,52){$\Lambda(\rho_{a|x})$}
\put(-72,25){$\rho_{AB}$}
\put(-70,-20){$(c)$}
\put(-5,8){$b$}
\put(-5,40){$y$}
\put(-130,8){$a$}
\put(-130,40){$x$}
\vspace{5mm}
  \caption{Depiction of quantum steering and device-independent protocols, where Alice and Bob share entangled state $\rho_{AB}$. The elements that generate the assemblage are depicted in purple and Bob's local operations on the normalized assemblage elements are illustrated in orange. (a) In the steering scenario, Alice performs uncharacterized measurements with settings $x$ and outcomes $a$, while Bob's subsystem is fully characterized. (b) Bob applies a local extraction channel -- a quantum operation on his subsystem that is independent of Alice's measurement choice. (c) In the device-independent protocol, Bob performs uncharacterized measurements on his subsystem $\rho_{a|x}$.
}
  \label{fig:1}
\end{figure}
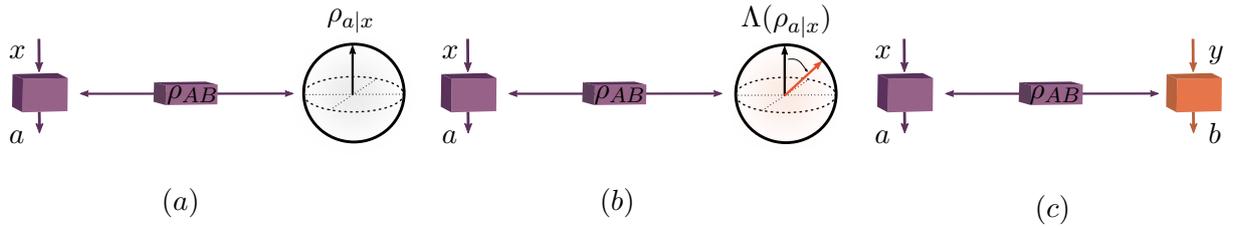

A steering scenario, illustrated in Fig.~\ref{fig:1}~(a), consists of two distant parties, Alice and Bob, who share a quantum system described by a state $\rho_{AB}\in \mathcal{L}(\cH_A \otimes\cH_B)$. Alice performs uncharacterized measurements: for each measurement setting $x \in \X$, she obtains an outcome $a \in \A$ with probability $p(a|x)$. The measurements are described by positive-operator-valued measurements (POVMs), denoted by $\{M_{a|x}\}_{a,x}$. Upon Alice's measurement, Bob's subsystem is described by a conditional state $\rho_{a|x}$. The main object of study in a steering scenario is the assemblage~\cite{pusey2013negativity}, defined as $\As_{\A|\X}=\{ \sigma_{a|x}\}_{a,x}$, where each unnormalized state $\sigma_{a|x}$ is given by $\sigma_{a|x} := p(a|x) \rho_{a|x}$. The assemblage elements all admit a quantum realization of the form
\begin{equation}
    \sigma_{a|x} = \tr{(M_{a|x} \otimes \id_B) \rho_{AB}}{A}.
\end{equation}
An assemblage is nonclassical if its elements cannot be decomposed as $\sigma_{a|x}=\sum_{\lambda} p(a|x\lambda)p(\lambda)\rho_\lambda$, i.e., if it cannot be simulated by Alice and Bob solely by local operations correlated by classical shared randomness. 

\vspace{5mm}

The nonclassicality of an assemblage can be detected using steering inequalities~\cite{cavalcanti2009experimental}. Any steering inequality can be written as
\begin{equation}\label{eq:steering-inequality}
    I[\As_{\A|\X}] = \Tr{\sum_{a \in \A,\, x \in \X} T_{ax} \, \sigma_{a|x}},
\end{equation}
where each $T_{ax}$ is a Hermitian operator. Let $\beta_C$ and $\beta_Q$ denote the classical and quantum bounds of the steering inequality, respectively. Then for any classical assemblage $\As^C_{\A|\X}$, we have $I[\As^C_{\A|\X}] \leq \beta_C$, whereas there exist nonclassical assemblages $\As_{\A|\X}$ such that $\beta_C < I[\As_{\A|\X}] \leq \beta_Q$.

To quantify the similarity between two assemblages, we use the concept of fidelity. Let $\As^{*}_{\A|\X}$ and $\As_{\A|\X}$ be two assemblages with the same input and output sets, and defined over the same Hilbert space $\cH_B$. We define the fidelity between them as 
\begin{equation}\label{eq:fidelity}
    \mathcal{F}(\As^{*}_{\A|\X}, \As_{\A|\X}) := \frac{1}{|\X|} \sum_{a,x} \sqrt{p^{*}(a|x) p(a|x)} \, F\left(\rho^{*}_{a|x}, \rho_{a|x}\right),
\end{equation}
where $F\left(\rho^{*}_{a|x}, \rho_{a|x}\right)$ is the Ulhmann-Josza fidelity between two quantum states\footnote{ This definition of fidelity between two assemblages is well-motivated for studying the distance between assemblages in the context of robust self-testing~\cite[Section 2]{Chen2021robustselftestingof}. There exist alternative definitions of fidelity between two assemblages, which preserve the typical properties of a fidelity-like quantity. For example, Ref.~\cite{nery2020distillation} uses a definition based on minimization over $x$ instead of averaging.}. When one of the states is pure, the fidelity reduces to $F(\rho^*_{a|x}, \rho_{a|x}) = \Tr{\rho^*_{a|x} \rho_{a|x}}$, and Eq.~\eqref{eq:fidelity} simplifies accordingly. A natural question then arises: given a nonclassical assemblage $\As_{\A|\X}$, what is its maximum fidelity with any classical assemblage? This quantity is given by the following optimization problem
\begin{equation}\label{eq:free-bound}
 \mathcal{F}^{C}(\As_{\A|\X}) = \mathop{\mathrm{max}}_{\As^{C}_{\A|\X}} \ \mathcal{F}(\As^{C}_{\A|\X}, \As_{\A|\X}),
\end{equation}
where the maximization is taken over all classical assemblages $\As^{C}_{\A|\X}$. Given $\As_{\A|\X}$, the fidelity $\mathcal{F}^{C}(\As_{\A|\X})$ can be computed with a semidefinite program presented in~\cite[Eq.~(12)]{Chen2021robustselftestingof}, which we recall in Appendix~\ref{app:freeCHSHsaturation}.

\subsection{Robust self-testing}\label{sec:robustST}

Consider a scenario where two parties, Alice and Bob, share an assemblage $\As_{\A|\X}$, with Bob's conditional states defined on a Hilbert space $\cH_B$ of unknown dimension. Since the exact form of $\As_{\A|\X}$ is not directly accessible, the goal in robust self-testing is to certify that this assemblage is similar to a known reference assemblage $\As^{ref}_{\A|\X}$. To quantify this similarity, we adopt the notion of \emph{extractability}~\cite{bardyn2009device,kaniewski2016analytic}, which formalizes the idea that two assemblages are equivalent if one can be transformed into the other by local operations. This concept captures the intuition that similarity between assemblages implies the existence of local channels that `extract' the reference assemblage from the underlying one. In this work, we quantify distance between assemblages using the fidelity introduced in Eq.~\eqref{eq:fidelity}, and we restrict our studies to local operations acting only on Bob’s subsystem. Formally, the extractability of $\As^{ref}_{\A|\X}$ form $\As_{\A|\X}$ is defined as
\begin{equation}\label{eq:E}
    \Xi(\As_{\A|\X} \rightarrow \As^{ref}_{\A|\X}) = \mathop{\mathrm{max}}_{\Lambda} \, \mathcal{F}(\As^{ref}_{\A|\X}\,, \Lambda(\As_{\A|\X})).
\end{equation}
Here, $\Lambda$ is a quantum channel applied to each assemblage element, i.e., $\Lambda(\As_{\A|\X})=\{\Lambda(\sigma_{a|x})\}$, which we refer to as an extraction channel. The concept of applying the extraction channel is illustrated in Fig.~\ref{fig:1}~(b).

In this paper, we focus on device-independent self-testing of quantum assemblages, where Bob's local dimension is unknown, and so are the measurements that he performs on his subsystem (see Fig.~\ref{fig:1}~(c)). Instead of analyzing the full correlations arising from Bob’s measurements, we focus on the value of a steering inequality, denoted $I[\As_{\A|\X}]=\beta$, which depends on both the assemblage elements and Bob's measurements. Hereon we assume that the reference assemblage achieves the maximal quantum violation of the inequality $I[\As^{ref}_{\A|\X}]=\beta_Q$. Self-testing based on steering inequalities is a common practice when considering one-sided device-independent self-testing of entangled states and measurements~\cite{supic2016,gheorghiu2017rigidity,goswami2018one,shrotriya2021robust,sarkar2023self}, where Bob is a trusted party. Here, we focus on the device-independent setting, where Bob's measurements are uncharacterized~\cite{Chen2021robustselftestingof}. 

In the case of perfect self-testing, extractability as defined in Eq.~\eqref{eq:E} equals one, which certifies that the underlying assemblage is equivalent to the reference assemblage. In robust self-testing, given a steering inequality violation $I[\As_{\A|\X}]=\beta$ with $\beta_C <\beta \leq \beta_Q$, we are interested in the minimum extractability over all assemblages that give this violation, formally defined as
\begin{equation}\label{eq:Q}
    Q_{I,\As^{ref}_{\A|\X}}(\beta) = \mathop{\mathrm{min}}_{\substack{\As_{\A|\X} \\ I[\As_{\A|\X}] = \beta}}  \mathop{\mathrm{max}}_{\Lambda}\,\,\, \mathcal{F}(\As^{ref}_{\A|\X}\,, \Lambda(\As_{\A|\X})).
\end{equation}

To place an upper bound on $Q_{I,\As^{ref}_{\A|\X}}(\beta)$, consider its value for two extreme cases. First, if $\beta = \beta_C$, the bound is trivial and given by $\mathcal{F}^{C}(\As^{ref}_{\A|\X})$. Second, the maximal violation $\beta = \beta_Q$ implies that the underlying assemblage is equivalent to the reference one, resulting in $Q_{I,\As^{ref}_{\A|\X}}(\beta_Q) = 1$. Any violation between the classical and quantum bounds can be obtained by a convex combination of assemblages achieving these extreme values. Consequently, the upper bound is given by
\begin{equation}\label{eq:Q-upper}
    Q_{I,\As^{ref}_{\A|\X}}(\beta) \leq \mathcal{F}^{C}(\As^{ref}_{\A|\X}) + (1-\mathcal{F}^{C}(\As^{ref}_{\A|\X})) \frac{\beta - \beta_C}{\beta_Q - \beta_C}.
\end{equation}
This upper bound is analogous to the one derived for robust self-testing of quantum states in Ref.~\cite{kaniewski2016analytic}. 

Deriving meaningful lower bounds for robust self-testing remains a central line of research in the field. Various approaches have been developed, including norm inequalities~\cite{mckague2012robust,mckague2011self,yang2013robust}, the numerical Swap method~\cite{bancal2015physical,yang2014robust}, and operator inequalities~\cite{kaniewski2016analytic}. In the next section, we derive analytic lower bounds on $Q_{I,\As^{\mathrm{ref}}_{\A|\X}}(\beta)$ for the CHSH-type assemblage.

\subsection{Comparative overview of self-testing statements}

As outlined in the introduction, self-testing is a broad framework applicable to a wide range of quantum resources. Although all self-testing statements rely on the same central idea -- namely, the unique certification of quantum resources -- it is useful to distinguish between different formulations and to emphasize the differences in their assumptions. In this subsection, we present a comparative overview of robust self-testing statements for different types of quantum resources: bipartite quantum states, quantum assemblages, and preparations in prepare-and-measure scenarios. This comparison clarifies how our results fit into the existing literature. We emphasize that we only review the results directly relevant to our work. For a comprehensive overview of self-testing, we refer the reader to Ref.~\cite{vsupic2020self}.

To formalize robust self-testing of different resources, we adopt the notion of \emph{extractability} defined in Section~\ref{sec:robustST}. Within a set of resources $\mathcal{R}$, it quantifies the similarity between a reference resource $R^{\mathrm{ref}} \in \mathcal{R}$ and the underlying resource, optimized over extraction channels $\Lambda^{\mathcal{R}}$. The notion of similarity, as well as the allowed extraction maps, depend on the specific type of resource and the assumptions of the self-testing scenario. Robust self-testing is then defined as follows: given an observed quantum correlation, minimize extractability over all compatible resources $R \in \mathcal{R}$ that reproduce this correlation. Let $D(R,R^{\mathrm{ref}})$ denote an appropriate distance measure and $I^{\mathcal{R}}[R]=\beta$ be the observed violation. Robust self-testing statement is then defined as follows
\begin{equation}\label{eq:Q-R}
Q_{I^{\mathcal{R}},R^{\mathrm{ref}}}(\beta) =
\min_{\substack{R \\ I^{\mathcal{R}}[R]=\beta}}
\max_{\Lambda^{\mathcal{R}}}\,
D\!\left(R^{\mathrm{ref}}, \Lambda^{\mathcal{R}}(R)\right).
\end{equation}

Table~\ref{tab:self-testing} summarizes robust self-testing of different quantum resources. For each scenario we specify: the type of resource, the reference resource, the notion of self-testing, the relevant observed correlation used to certify nonclassicality of the resource, the admissible extraction channels, and the measure adopted to quantify similarity.

Bipartite quantum states can be self-tested via a Bell scenario~\cite{MY98,MY04,Bell-Nonlocality,vsupic2020self,kaniewski2016analytic}, or a steering scenario~\cite{supic2016,gheorghiu2017rigidity}. In a Bell scenario, we consider two parties, Alice and Bob, who share a bipartite system prepared in a state $\rho \in \mathcal{L}(\mathcal{H}_A \otimes \mathcal{H}_B)$. Alice and Bob perform POVMs, denoted $\{M_{a|x}\}$ and $\{M_{b|y}\}$, which give rise to the observed correlations $p(ab|xy) = \tr{(M_{a|x} \otimes M_{b|y})\rho}{}$. To derive a robust self-testing statement, rather than analyzing these probabilities directly, one may instead consider the violation of a Bell inequality, defined as
$\mathcal{B}\,[\,p(ab|xy)\,] = \sum_{a,b,x,y} \alpha_{abxy}\, p(ab|xy)$. In this setting, the extraction channels are given by local operations on each party and the similarity can be measured with vector norm, fidelity or trace-distance~\cite{vsupic2020self}. The central assumption in the Bell scenario is the no-signaling principle. Under this condition, one can consider robust self-testing as a device-independent protocol, in which Alice’s and Bob’s devices are uncharacterized, 
and any conclusions about the shared quantum state follow solely from the observed correlations $p(ab|xy)$. Moreover, in device-independent self-testing, no assumption is made about the dimension of the underlying quantum system.

To relax the assumption of device-independence, bipartite quantum states can be self-tested within a steering scenario~\cite{supic2016}. In an assemblage-based one-sided self-testing, Bob is a trusted party who can fully characterize the assemblage $\As_{\A|\X}$ as well as his reduced state $\rho_B=\tr{\rho}{A}$. The task of robust self-testing is then to extract the information about the shared state from these two resources. This assumption can be relaxed by allowing Bob to perform characterized measurements without granting him 
the ability to carry out full tomography of his local subsystem. In this correlation-based one-sided approach, self-testing is based on the violation of an EPR steering inequality, 
which depends on the underlying assemblage together with Bob’s characterized measurements. 
In both formulations described above, the extraction channels are local operations on Alice's side, as Bob is a trusted party. To measure similarity between resources, Ref.~\cite{supic2016} uses trace distance for quantum states and Schatten 1-norm for assemblage elements.

The scenario investigated in this work is the robust self-testing of assemblages~\cite{Chen2021robustselftestingof}, 
which is discussed in detail in Section~\ref{sec:robustST}. As summarized in Table~\ref{tab:self-testing}, the relevant resource is not the underlying bipartite state, 
but the collection of unnormalized states that constitute the assemblage. 
The notion is device-independent: both parties are uncharacterized, Bob’s measurements are unknown, and no restriction is imposed on the dimension of his system. The observed correlation is captured by the violation of an EPR steering inequality. Since the self-tested systems are local to Bob, the extraction channels are restricted to local operations on his side. Similarity is measured with assemblage fidelity.

The final scenario considered in Table~\ref{tab:self-testing} is the prepare-and-measure setting~\cite{tavakoli2018self}. Here, a preparation device generates quantum states $\rho_x \in \mathcal{L}(\mathcal{H}_A)$ depending on a classical input $x \in \mathcal{X}$, 
which are subsequently measured with POVMs $\{M_{b|y}\}$. 
The statistics of the experiment are given by $p(b|x,y) = \operatorname{Tr}[M_{b|y} \rho_x]$. Nonclassicality in this scenario can be certified using dimension witnesses of the form $\mathcal{A}\,[\,p(b|x,y)\,] = \sum_{x,y,b} \alpha_{xyb}\, p(b|x,y)$. 
The admissible extraction channels act solely on Alice's Hilbert space and similarity between two sets of quantum states is assessed with average fidelity.
This setting differs from the previously discussed ones in that it requires an assumption on the dimension of the prepared systems. Because of this assumption, it is referred to as semi-device-independent.

\begin{table}[h!]
\centering
\caption{Overview of robust self-testing for different quantum resources.}
\label{tab:self-testing}
\scriptsize
\renewcommand{\arraystretch}{1.3}
\begin{tabular}{|c|c|c|c|c|c|}
\hline
\shortstack{\textbf{Resource}} &
\shortstack{\textbf{Reference} \\ \textbf{resource}} &
\shortstack{\textbf{Notion}} &
\shortstack{\textbf{Observed} \\ \textbf{correlation}} &
\shortstack{\textbf{Extraction} \\ \textbf{channels}} &
\shortstack{\textbf{Similarity} \\ \textbf{measure}} \\
\hline
\shortstack{Bipartite \\ quantum states} &
\shortstack{$\rho^{\mathrm{ref}} \in \mathcal{L}(\mathcal{H}_A \otimes \mathcal{H}_B)$} &
\shortstack{Device-independent \\ \, \\ \,} &
\shortstack{$\mathcal{B}[p(ab|xy)] = \beta$ \\ \,\\ \,} &
\shortstack{$\Lambda_A \otimes \Lambda_B$ \\ \,\\ \,} &
\shortstack{Vector norm, \\ fidelity, \\ trace distance} \\
\cline{3-6}
 & & \shortstack{Assemblage-based \\ one-sided} &
\shortstack{$\rho_B,\, \As_{\A|\X}$} &
\shortstack{$\Lambda_A$} &
\shortstack{Trace distance, \\ Schatten 1-norm} \\
\cline{3-6}
 & & \shortstack{Correlation-based \\ one-sided} &
\shortstack{$I[\As_{\A|\X}] = \beta$} &
\shortstack{$\Lambda_A$} &
\shortstack{Trace distance, \\ Schatten 1-norm} \\
\hline
\shortstack{Quantum \\ assemblages} &
\shortstack{$\As^{\mathrm{ref}}_{\A|\X}$ \\ with $\sigma^{\mathrm{ref}}_{a|x} \in \mathcal{L}(\mathcal{H}_B)$} &
\shortstack{Device-independent\\ \,} &
\shortstack{$I[\As_{\A|\X}] = \beta$\\ \,} &
\shortstack{$\Lambda_B$\\ \,} &
\shortstack{Assemblage fidelity\\ \,} \\
\hline
\shortstack{Prepare-and- \\ measure states} &
\shortstack{$\{\rho^{\mathrm{ref}}_x\}_{x \in \X}$ \\ with $\rho^{\mathrm{ref}}_x \in \mathcal{L}(\mathcal{H}_A)$} &
\shortstack{Semi-device-independent\\ \,} &
\shortstack{$\mathcal{A}[p(b|x,y)] = \beta$\\ \,} &
\shortstack{$\Lambda_A$\\ \,} &
\shortstack{Average fidelity\\ \,} \\
\hline
\end{tabular}
\end{table}

\medskip

Finally, we remark that the notion of extractability is naturally connected to self-testing in the framework of resource theories~\cite{schmid2020standard}. In this language, the extractability can be understood as specifying possible resource conversions, where the free operations are given by the set of possible extraction channels.

\section{Results}

Our main result provides an analytic lower bound for device-independent self-testing of the CHSH-type assemblage as a function of the observed CHSH violation. We begin by defining this reference assemblage and then construct the necessary operator inequalities.

We denote the CHSH-type assemblage by $\As^{*}_{\A|\X}=\{\sigma^{*}_{a|x} = p^{*}(a|x)\rho^{*}_{a|x}\}$, where the unnormalized states are defined as follows:
\begin{equation}\label{eq:CHSH-assemblage}
\begin{aligned}
    \sigma^{*}_{0|0}& =\frac{1}{2} \ket{0}\bra{0}, \quad
    \sigma^{*}_{1|0} = \frac{1}{2} \ket{1}\bra{1}, \\
    \sigma^{*}_{0|1} &= \frac{1}{2} \ket{+}\bra{+}, \quad 
    \sigma^{*}_{1|1} = \frac{1}{2} \ket{-}\bra{-}.
\end{aligned}
\end{equation}
One possible realization of this assemblage involves Alice and Bob sharing the maximally entangled two-qubit state $\ket{\phi^{+}}=\frac{1}{\sqrt{2}}(\ket{00}+\ket{11})$, and Alice applying Pauli observables $\{A_0,A_1\}=\{Z,X\}$. To construct a robust self-test of $\As^{*}_{\A|\X}$, we use the CHSH inequality adapted to the steering scenario:
\begin{align}\label{eq:CHSH-inequality}
I_{CHSH} &= \av{A_0B_0} + \av{A_0B_1} + \av{A_1B_0} - \av{A_1B_1} \nonumber \\
&= \Tr{(B_0+B_1) \, \sigma_{0|0} - (B_0+B_1) \, \sigma_{1|0} + (B_0-B_1) \, \sigma_{0|1} - (B_0-B_1) \, \sigma_{1|1}  }. 
\end{align}
This expansion of $I_{CHSH}$ naturally defines a steering inequality of the form given in Eq.~\eqref{eq:steering-inequality} with the operators $T_{00} = B_0 + B_1 = -T_{10}$ and $T_{01} = B_0 - B_1 = -T_{11}$. Here, $B_0$ and $B_1$ are Bob's observables. Since we focus exclusively on the CHSH inequality, we consider the case where Bob has two dichotomic measurements. In this setting, Jordan’s lemma~\cite[Lemma 2]{pironio2009device} allows us to reduce Bob's subsystem to a qubit and express his observables as $B_0=\cos(\theta) Z + \sin (\theta)X$ and $B_1=\cos(\theta) Z - \sin (\theta)X$, with $\theta \in [0,\frac{\pi}{2}]$.

Let $\As_{\A|\X}=\{\sigma_{a|x} = p(a|x)\rho_{a|x}\}$ be a valid quantum assemblage. Hereon, we always assume the sets $\A$ and $\X$ are the same for $\As_{\A|\X}$ and the reference assemblage $\As^{*}_{\A|\X}$. We consider the extractability of the CHSH-type assemblage, defined as
\begin{equation}
    \Xi(\As_{\A|\X} \rightarrow \As^{*}_{\A|\X}) = \mathop{\mathrm{max}}_{\Lambda} \,\,\mathcal{F}(\As^{*}_{\A|\X}, \Lambda(\As_{\A|\X})).
\end{equation}
Inserting the definition of fidelity from Eq.~\eqref{eq:fidelity}, we obtain
\begin{equation}\label{eq:Q-CHSH}
    \Xi(\As_{\A|\X} \rightarrow \As^{*}_{\A|\X}) = \mathop{\mathrm{max}}_{\Lambda} \frac{1}{|\X|} \sum_{a,x} \sqrt{\frac{p^{*}(a|x)}{p(a|x)}} \, \Tr{\rho^{*}_{a|x}\Lambda( \sigma_{a|x})}.
\end{equation}
As the maximum violation of the steering inequality defined in Eq.~\eqref{eq:CHSH-inequality} is only achievable when $p^{*}(a|x)=p(a|x)$~\cite{goh2018geometry}, we assume this equality holds throughout the derivation below\footnote{Note that the results of Ref.~\cite{Chen2021robustselftestingof} are also derived under this assumption; therefore, this does not limit the comparison of our analytic bounds to their numerical results.}. Define the operators
\begin{equation}
    K_{ax} = \Lambda^{\dagger} (\rho^{*}_{a|x})
\end{equation}
for each $x$ and $a$, where $\Lambda^{\dagger}$ denotes the dual of the quantum channel $\Lambda$. Note that since $\Lambda$ is not fixed in the optimization problem, the operators $K_{ax}$ generally depend on the specific choice of the observables $B_0$ and $B_1$. Consequently, we obtain the following lower bound on extractability:
\begin{equation}
    \Xi(\As_{\A|\X} \rightarrow \As^{*}_{\A|\X})  \geq \frac{1}{|\X|} \sum_{a,x} \Tr{K_{ax}\,\sigma_{a|x}}.
\end{equation}

Our goal is to construct an operator inequality of the form 
\begin{equation}
    K_{ax} \geq s \,T_{ax} + t_{ax} \,\id,
\end{equation}
which holds for all assemblage elements -- that is, for every pair $(a,x)$ -- and for all possible choices of Bob's measurements. Here, $s$ and $t_{ax}$ are real coefficients to be appropriately chosen later. Evaluating this operator inequality on a given assemblage $\As_{\A|\X}$ and taking the trace yields the following bound:
\begin{equation}
    \frac{1}{|\X|} \sum_{a,x} \Tr{K_{ax}\,\sigma_{a|x}} \\
    \geq \frac{s}{|\X|}  \Tr{\sum_{a,x} T_{ax}\,\sigma_{a|x}} + \frac{1}{|\X|}\sum_{a,x}\,p(a|x)\,t_{ax}. 
\end{equation}
For the choice of $T_{ax}$ corresponding to the steering inequality defined in Eq.~\eqref{eq:CHSH-inequality}, this leads to a lower bound on extractability:
\begin{equation}\label{eq:CHSH-extr-bound}
    \Xi(\As_{\A|\X} \rightarrow \As^{*}_{\A|\X}) 
    \geq \frac{s}{|\X|}\, I_{CHSH}[\As_{\A|\X}] +  \frac{1}{|\X|}\sum_{a,x}\,p(a|x)\,t_{ax}. 
\end{equation}

\begin{figure}[ht!]
  \centering
  \includegraphics[width=0.8\linewidth]{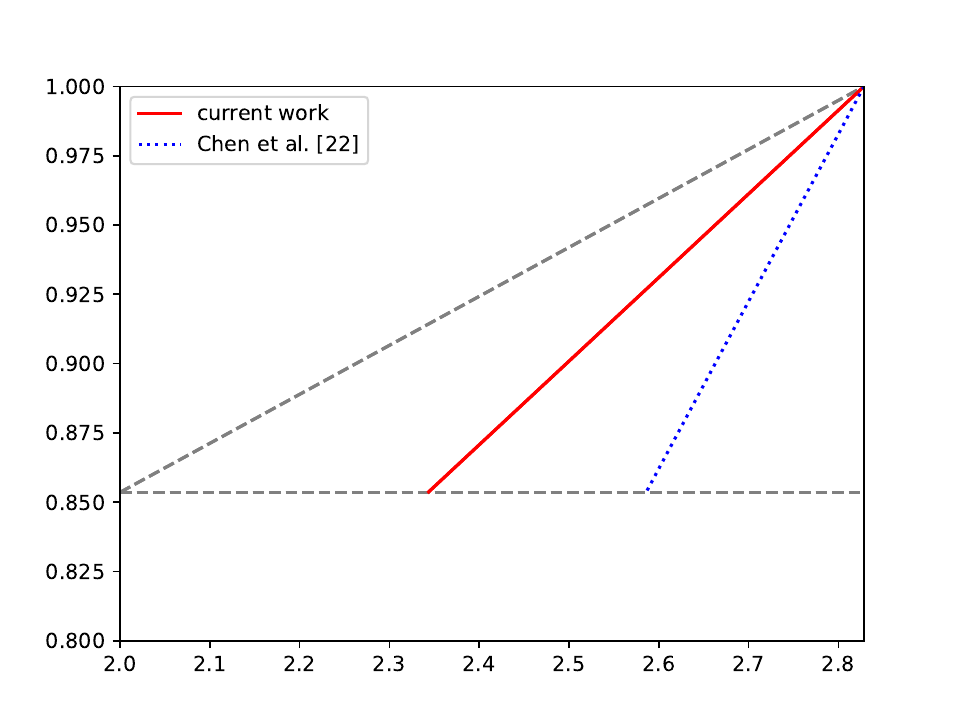}
\put(-390,110){\rotatebox{90}{$Q_{I_{CHSH},\As^{*}_{\A|\X}}(\beta)$}}
\put(-180,0){$\beta$}
  \caption{Robust self-testing bound for the CHSH-type assemblage. The red line corresponds to the lower bound on $Q_{I_{CHSH},\As^{*}_{\A|\X}}(\beta)$ derived in this paper. The dotted blue line shows the numerical bound obtained in Ref.~\cite{Chen2021robustselftestingof}. The gray dashed horizontal line corresponds to the trivial threshold $\mathcal{F}^{C}(\As^{*}_{\A|\X})=(2+\sqrt{2})/4$, which can be attained by a classical assemblage. The upper bound given in Eq.~\eqref{eq:Q-upper} is also represented by a gray dashed line.}
  \label{fig:chsh_plot}
\end{figure}

To obtain a robust self-testing statement, we aim to lower bound $Q_{I_{CHSH},\As^{*}_{\A|\X}}(\beta)$, which involves a minimization over all assemblages $\As_{\A|\X}$ and observables $B_0,B_1$ satisfying $I_{CHSH}[\As_{\A|\X}]=\beta$. Applying this minimization to the right-hand side of Eq.~\eqref{eq:CHSH-extr-bound}, assuming uniform probabilities $p(a|x) = \frac{1}{2}$, we obtain
\begin{equation}
    Q_{I_{CHSH},\As^{*}_{\A|\X}}(\beta) \geq \frac{1}{2} ( s \beta + \frac{1}{2}\min_{B_0,B_1}t_{ax}) .
\end{equation}
In Appendix~\ref{app:CHSHbound}, we provide an explicit construction of the channel $\Lambda$ that yields the lower bound specified by the coefficients:
\begin{equation}\label{eq:chsh-coeff}
    s =\frac{\sqrt{2}+1}{4}, \qquad t = \frac{2-\sqrt{2}}{2},
\end{equation}
where $t \equiv\frac{1}{2}\min_{B_0,B_1}t_{ax}$. This result provides a robust self-testing statement for the CHSH-type assemblage. Our analytic bound on the fidelity with the CHSH-type assemblage significantly improves the previously known numerical lower bound from Ref.~\cite{Chen2021robustselftestingof}, which was derived using a semidefinite program. The comparison between the bounds is illustrated in Figure~\ref{fig:chsh_plot}. 

The bound specified by coefficients given in Eq.~\eqref{eq:chsh-coeff} can be used to determine the threshold violation above which the bound is non-trivial. The trivial bound for the CHSH-type assemblage, given by $\mathcal{F}^{C}(\As^{*}_{\A|\X})$, was computed in Ref.~\cite{Chen2021robustselftestingof} using a semidefinite program and found to be approximately $0.85355$. This value is depicted in Figure~\ref{fig:chsh_plot} as a gray horizontal line. In Appendix~\ref{app:freeCHSHsaturation}, we present a strategy that achieves the value $(2+\sqrt{2})/4$, thereby saturating this lower bound. By examining the intersection of the trivial bound with the lower bound derived in this paper, it follows that the threshold violation is equal to $8 - 4\sqrt{2} \approx 2.34$. This result significantly improves on the threshold for obtaining non-trivial fidelity with the CHSH-assemblage from Ref.~\cite{Chen2021robustselftestingof}, which is given by $4-\sqrt{2} \approx 2.59$.

\section{Conclusions and outlook}

We have derived the first analytic bound for device-independent self-testing of steering assemblages. By successfully adapting the operator inequalities framework to the steering scenario, we obtained explicit robustness bounds for the CHSH-type assemblage that significantly improve upon existing numerical results. This extension of analytic self-testing methods from quantum states and measurements to assemblages demonstrates the generality of the operator inequalities approach and its potential for broader applications.

Our results are currently limited to the CHSH-type assemblage that plays a crucial role in many quantum information processing protocols. A natural direction for future work is to extend these methods to other assemblages. However, constructing optimal extraction channels remains technically challenging, particularly when the outcome probabilities $p(a|x)$ are not uniform across all assemblage elements. While our work does not resolve this general issue, it demonstrates that the operator inequalities framework continues to provide meaningful bounds worth pursuing in future research.

Our analytic bound open several research directions. From a theoretical perspective, explicit self-testing bounds could facilitate quantitative studies of the relationship between steering assemblages and the boundary of the set of quantum correlations. From an experimental perspective, an interesting direction would be developing device-independent steering measures that leverage analytic robustness bounds, building on recent experimental work such as Ref.~\cite{zhao2020experimental}.

Several open problems remain inherent to the framework of robust self-testing via operator inequalities, making generalization to other types of steering assemblages challenging. Since the approach relies on Jordan's lemma, it is currently applicable only to measurements with two inputs and two outputs; extending beyond this would allow the study of different self-testing inequalities, such as the three-input scenarios discussed in Ref.~\cite{Wang2022}. Additionally, while there has been work on self-testing complex-valued assemblages~\cite{Chen2021robustselftestingof}, these methods rely on modifying the definition of assemblage fidelity, which poses a challenge for adapting the techniques used in this paper. Another possible direction is to extend the operator inequality method to self-testing of high-dimensional assemblages. Although some inequalities have been proposed for certifying genuine high-dimensional steering~\cite{Designolle2021,chang2023experimental}, deriving analytical robustness bounds for qudit self-testing remains a significant challenge, and applying operator inequalities in this context could provide a powerful approach for future research.
 
In this paper, we considered a general scenario in which the underlying assemblage is unrestricted, allowing Alice to perform arbitrary POVM measurements. Mutually unbiased basis (MUB) measurements are known to be a valuable resource for quantum steering~\cite{Marciniak2015,Jebarathinam2018,Zhu2016,Bavaresco2017}, and one could restrict the assemblages to those arising from MUB measurements to study how the robustness bounds are affected. This could be relevant in practical scenarios where the assemblage is unknown, but it is possible to certify that Alice’s measurements form MUBs. However, this problem is interesting in dimensions where MUB sets are not unitarily equivalent (such as $d=4$~\cite{mub}), which poses a challenge to the techniques applied in this work. Moreover, it is worth noting that steering has been used to study self-testing of MUBs in Ref.~\cite{Sarkar2022}. It would be interesting to investigate whether the operator inequalities approach can be extended to this problem.

The main motivation for studying robust self-testing bounds is to enable resource certification in realistic experiments affected by noise. While self-testing of assemblages is a recent concept, self-testing of states has already been demonstrated across several platforms. Although self-testing of assemblages requires stronger Bell-inequality violations than state self-testing to achieve nontrivial fidelities, current results suggest this is within reach. Device-independent self-testing has been realized in photonic systems~\cite{Zhang2018,Zhang2019}, trapped ions~\cite{Tan2017}, and superconducting qubits~\cite{Wu2022,Storz2025}, achieving high fidelities. These results indicate that extending self-testing protocols to assemblages is experimentally feasible with near-term technologies, providing a clear pathway for future demonstrations.

\section*{Acknowledgments}
BZ thanks Anubhav Chaturvedi for useful discussions. BZ acknowledges support by the National Science Centre, Poland 2021/41/N/ST2/02242. 

\bibliography{main-review}
\bibliographystyle{unsrtnat}

\appendix

\section{Operator inequalities for self-testing assemblages}

\label{app:CHSHbound}

In this Appendix, we provide a derivation of the lower bound on the fidelity with the CHSH-type assemblage specified by coefficients given in Eq.~\eqref{eq:chsh-coeff}. This result utilizes the method of operator inequalities which was first introduced in Ref.~\cite{kaniewski2016analytic} for robust-self testing of a maximally entangled state. The proof we present here is more similar to the one in Ref.~\cite{tavakoli2018self}, where operator inequalities are used to bound robust self-testing result for a prepare and measure scenario.

In the main text, we denote the reference assemblage as $\As^{*}_{\A|\X}=\{\sigma^{*}_{a|x} = p^{*}(a|x)\rho^{*}_{a|x}\}$. Its elements can be written as
\begin{equation}
\begin{aligned}
    \sigma^{*}_{0|0} &=\frac{1}{2} \ket{0}\bra{0}, \quad
    \sigma^{*}_{1|0} = \frac{1}{2} \ket{1}\bra{1}, \\
    \sigma^{*}_{0|1} &= \frac{1}{2} \ket{+}\bra{+}, \quad 
    \sigma^{*}_{1|1} = \frac{1}{2} \ket{-}\bra{-}.
\end{aligned}
\end{equation}
It is easy to see that each normalized state $\rho^{*}_{a|x}$ is assigned probability $p^{*}(a|x)=\frac{1}{2}$.

Recall that our goal is to construct an operator inequality of the form 
\begin{equation}\label{eq:operator-in-CHSH}
    K_{ax} \geq s \,T_{ax} + t_{ax} \,\id.
\end{equation}
The operators $K_{ax}$ are given by
\begin{equation}
    K_{ax} = \Lambda^{\dagger} (\rho^{*}_{a|x}),
\end{equation}
where $\rho^{*}_{a|x}$ are the normalized states of the reference assemblages. We take the channel $\Lambda$ to be the following
\begin{equation}
\Lambda_\theta(\rho) = \frac{1 + c(\theta)}{2} \rho + \frac{1 - c(\theta)}{2} \Gamma(\theta) \rho \Gamma(\theta), 
\end{equation}
where for $0 \leq \theta \leq \pi/4$ we use $\Gamma = Z$, while for $\pi/4 < \theta \leq \pi/2$ we use $\Gamma = X$. The function $c(\theta) \in [-1,1]$ will be specified later. Notice that this channel is self-dual. For $0 \leq \theta \leq \pi/4$, the channel applied on the normalized state of Bob's reference assemblage leads to
\begin{equation}
\begin{aligned}
K_{00} &= \frac{\id + Z}{2}, \quad && K_{01} = \frac{\id + c(\theta) X}{2}, \\
K_{10} &= \frac{\id - Z}{2}, \quad && K_{11} = \frac{\id - c(\theta) X}{2},
\end{aligned} 
\end{equation}
whereas in the interval $\pi/4 < \theta \leq \pi/2$, we have
\begin{equation}
\begin{aligned}
K_{00} &= \frac{\id + c(\theta) Z}{2}, \quad && K_{01} = \frac{\id + X}{2}, \\
K_{10} &= \frac{\id - c(\theta) Z}{2}, \quad && K_{11} = \frac{\id - X}{2}.
\end{aligned} 
\end{equation}
As explained in the main next, we use the Jordan’s lemma to express Bob's measurements as 
\begin{equation}
    B_0 = \cos(\theta) Z + \sin(\theta)X, \quad B_1 = \cos(\theta) Z - \sin(\theta)X.
\end{equation}
This parametrization covers all possible choices of Bob's observables and ensures the operator inequality holds for arbitrary observables, not only qubit ones. Recall that the operators $T_{ax}$ defining the steering inequality given in Eq.~\eqref{eq:CHSH-inequality} are given by $T_{00} = B_0 + B_1 = -T_{10}, T_{01} = B_0 - B_1 = -T_{11}$; hence, they can be written as
\begin{equation}
    T_{00} = - T_{10} = 2\cos(\theta) Z, \quad T_{01} = - T_{11} = 2\sin(\theta) X.
\end{equation}
Notice that the operators $K_{ax}$ and $T_{ax}$ labeled by $(a=0, x=0)$ and $(a=1, x=0)$ differ only by the sign of the operator $Z$, and the operators labeled by $(a=0, x=1)$ and $(a=1, x=1)$ differ only by the sign of the operator $X$. We can therefore reduce the number of inequalities by defining $t_{0}\equiv t_{00}=t_{10} $ and $t_{1}\equiv t_{01}=t_{11}$. We can now explicitly write the operator inequalities of Eq.~\eqref{eq:operator-in-CHSH}. In the first interval, we have
\begin{equation}\label{eq:app1}
\frac{\id + Z}{2} - 2s \cos\theta\, Z - t_0\, \id \geq 0, \quad
\frac{\id + c(\theta) X}{2} - 2s \sin\theta\, X - t_1\, \id \geq 0.
\end{equation}
In the second interval, the two operator inequalities are
\begin{equation}\label{eq:app2}
\frac{\id + c(\theta) Z}{2} - 2s \cos\theta\, Z - t_0\, \id \geq 0, \quad
\frac{\id + X}{2} - 2s \sin\theta\, X - t_1\, \id \geq 0. 
\end{equation}
For the operators given in Eqs.~\eqref{eq:app1} and \eqref{eq:app2} to be positive semi-definite in all regions, $t_0$ and $t_1$ need to satisfy the constraints implied by both intervals. Solving Eqs.~\eqref{eq:app1} and \eqref{eq:app2} leads to two constraints for $t_0$ and $t_1$ in each interval. Any choice of $t_0$ and $t_1$ satisfying these constraints gives rise to valid operator inequalities. In order to obtain the strongest bound, we minimize the possible values of $t_0$ and $t_1$ over all possible measurement operators, which correspond to different values of the angle $\theta$. Then, in the first interval
\begin{equation}\label{eq:t1}
\begin{aligned}
t_0 &= \min\left\{ 1 - 2 s \cos\theta, 2 s \cos\theta \right\}, \\
t_1 &= \min\left\{ \frac{1}{2} (1 + c(\theta) - 4 s \sin\theta), \frac{1}{2} (1 - c(\theta) + 4 s \sin\theta) \right\}.
\end{aligned} 
\end{equation}
A similar procedure for the second interval leads to
\begin{equation}\label{eq:t2}
\begin{aligned}
t_0 &= \min\left\{ \frac{1}{2} (1 + c(\theta) - 4 s \cos\theta), \frac{1}{2} (1 - c(\theta) + 4 s \cos\theta) \right\}, \\
t_1 &= \min\left\{ 1 - 2 s \sin\theta, 2 s \sin\theta \right\}.
\end{aligned} 
\end{equation}
In the main text, we discuss how to use this technique to obtain the following bound on extractability
\begin{equation}\label{eq:CHSH-extr-bound-}
    \Xi(\As_{\A|\X} \rightarrow \As^{*}_{\A|\X}) 
    \geq \frac{s}{|\X|}\, I_{CHSH}[\As_{\A|\X}] +  \frac{1}{2 |\X|}\sum_{a,x}\,\,t_{ax}. 
\end{equation}
Define $t \equiv \frac{1}{2}\sum_{a,x}\,\,t_{ax}=t_0+t_1$. To obtain the lower bound, we choose the dephasing function as \( c(\theta) = \min\{1, 4 s \sin\theta\} \) for \( \theta \in [0, \pi/4] \), and \( c(\theta) = \min\{1, 4 s \cos\theta\} \) for \( \theta \in (\pi/4, \pi/2] \). Using the constraint on $t$ given in Eqs.~\eqref{eq:t1} and ~\eqref{eq:t2}, we then calculate the optimal lower bound on extractability by minimizing over $t$ for different values of $s$. We find the optimal lower bound corresponds to
\begin{equation}
s = \frac{1 + \sqrt{2}}{4}, \qquad t = \frac{2 - \sqrt{2}}{2}.
\end{equation}
It follows that
\begin{equation}
    Q_{I_{CHSH},\As^{ref}_{\A|\X}}(\beta) \geq   \frac{1 + \sqrt{2}}{8} \beta + \frac{2 - \sqrt{2}}{4} .
\end{equation}
Calculations presented in this Appendix together with a visualization of how the self-testing bound changes for different values of the parameter $s$ are available in an online repository~\cite{github-ST}.

\section{Trivial fidelity with the CHSH-type assemblage}
\label{app:freeCHSHsaturation}

Given a nonclassical assemblage $\As^{*}_{\A|\X}$, its maximum fidelity with any classical assemblage is defined as
\begin{equation}
 \mathcal{F}^{C}(\As^{*}_{\A|\X}) = \mathop{\mathrm{max}}_{\As^{C}_{\A|\X}}\ \mathcal{F}(\As^{C}_{\A|\X}, \As^{*}_{\A|\X}),
\end{equation}
where the maximization is over the set of classical assemblages. Recall that elements of classical assemblages admit a decomposition of the form $\sigma^{C}_{a|x}=\sum_{\lambda} p(\lambda) \delta_{a,\lambda_x} \rho_{\lambda}$, where $\delta_{a,\lambda_x}$ is a deterministic probability distribution and each $\rho_\lambda$ is a normalized quantum state. In Ref.~\cite{Chen2021robustselftestingof}, it was showed that $\mathcal{F}^{C}(\As^{*}_{\A|\X})$ for any nonclassical assemblage can be computed with the following semi-definite program:
\begin{equation}\label{SDP-trivial}
\max_{\{\rho_\lambda\}} \frac{\sqrt{|\A|}}{|\X||\lambda|} \sum_{a,x,\lambda} \sqrt{p^{*}(a|x)} \, \delta_{a,\lambda_x} \, \mathrm{tr}(\rho^{*}_{a|x} \, \rho_\lambda)
\quad \text{s.t.} \quad \rho_\lambda \geq 0, \quad \mathrm{tr}(\rho_\lambda) = 1 \quad \forall \lambda.
\end{equation}
For the CHSH-type assemblage, $\mathcal{F}^{C}(\As^{*}_{\A|\X})=0.8536$. 

Here, we present an explicit example of a classical assemblage that saturates this bound. It is generated in the following way. First, a classical variable $\lambda \in \{0,1\}$ with $p(\lambda=0)=p(\lambda=1)=\frac{1}{2}$ is distributed to Alice and Bob. Alice simply copies the value of $\lambda$ outputs it as her output $a$, ignoring the value of $x$. Bob on the other side outputs one of the following two states $\rho_\lambda$ depending on the value of $\lambda$:

\begin{equation}
\rho_1 = \frac{1}{2} \begin{bmatrix}
1 + \frac{\sqrt{2}}{2} & \frac{\sqrt{2}}{2} \\
\frac{\sqrt{2}}{2} & 1 - \frac{\sqrt{2}}{2}
\end{bmatrix},
\qquad
\rho_2 = \frac{1}{2} \begin{bmatrix}
1 - \frac{\sqrt{2}}{2} & -\frac{\sqrt{2}}{2} \\
-\frac{\sqrt{2}}{2} & 1 + \frac{\sqrt{2}}{2}\\
\end{bmatrix}.
\end{equation}
Keeping in mind that $p^{*}(a|x)=1/2$ for all $a$ and $x$, it is easy to see that fidelity between the classical assemblage constructed above and the CHSH-type assemblage is the following

\begin{equation}
\mathcal{F}^{C}(\As^{*}_{\A|\X}) = \frac{1}{4}(\Tr{\rho^{*}_{0|0} \, \rho_0}+\Tr{\rho^{*}_{1|0} \, \rho_1}+\Tr{\rho^{*}_{0|1} \, \rho_0}+\Tr{\rho^{*}_{1|1} \, \rho_1}) = \frac{2+\sqrt{2}}{4} \approx 0.85355.
\end{equation}

\end{document}